# Unravelling the role of vacancies in lead halide perovskite through electrical switching of photoluminescence

Cheng Li,[1] Antonio Guerrero,[2] Sven Huettner,[1] Juan Bisquert*[2]

[1]Department of Chemistry, University of Bayreuth, Universitätstr. 30, 95447 Bayreuth, Germany.
[2]Institute of Advanced Materials (INAM), Universitat Jaume I, 12006 Castello, Spain.

Email: bisquert@uji.es

Methylammonium lead triiodide perovskite (MAPbI$_3$) semiconductor displays outstanding photovoltaic and light emitting properties. We address the unique behavior in which a bias voltage can be used to switch on and off the luminescence of a planar film with lateral symmetric electrodes. Instead of a homogeneous suppression of emission, as in other organic semiconductor films, in MAPbI$_3$ films a dark region advances from the positive electrode at a slow velocity of order of 1μm s$^{-1}$. Here we explain the existence of the sharp front in terms of the drift of ionic vacancies that drastically reduce the radiative recombination rate in the film. Based on a dynamic transport model we show that the square reciprocal of the electrical current is linear with time in agreement with the experimental observations. This insight leads to a direct determination of the diffusion coefficient of iodine vacancies $D_{V_I} = 6 \times 10^{-9} \, \text{cm}^2 \text{s}^{-1}$ and provides detailed information and control on the effect of ionic conduction over the electrooptical properties of hybrid perovskite materials.

The advent of hybrid perovskite solar cells has given rise to extraordinary photovoltaic performances, causing the rise of a new solar energy conversion technology. However, the new physical characteristics of these materials are not yet completely understood, and many significant experimental observations have not been satisfactorily explained so far. Slow time scale variations of photoluminescence (PL) phenomena have been widely reported since an early stage in the research of perovskite photovoltaics and optoelectronic.[1,2] Sanchez et al. [3] observed either increasing or decreasing PL intensity according to the preparation method of the methylammonium lead triiodide (MAPbI$_3$) layer. Hoke et al. [4] interpreted the observation of light-induced transformations of PL in mixed MAPb(Br$_x$I$_{1-x}$)$_3$ in terms of photoinduced phase segregation. Subsequently, Leitjens et al. [5] observed that an applied electrical field across the perovskite layer can either enhance or suppress the luminescence in lateral interdigitated electrode devices. These findings were described in terms of a simple mechanism common to both fully inorganic and organic semiconductors [6,7] in which photogenerated electrons and holes drift to opposite sides of the device, reducing the bulk recombination rate and hence PL intensity. However, some aspects of the observations cannot be explained by this simple model. Furthermore it is expected that under an applied field a massive ion drift will strongly affect the PL characteristics. In fact the ion displacement has been reported to modify PL properties under different conditions [8,9] and a variety of reversible and irreversible PL transient responses have been obtained under modification of the applied bias magnitude and electrode polarity of laterally contacted MAPbI$_3$ layers[10-12] that still require a general explanation.

Recently a new property was uncovered, revealing striking features of the transient PL in perovskite layers.[13] Using a wide-field PL imaging microscope it was observed that the PL is progressively suppressed from the positively biased electrode. New data has been recorded for interdigitated electrodes with a wider channel length of ~150 µm and they are shown in Fig. 1a. The darkened area in the left of the image forms a sharp front that advances at a slow velocity of ca. 5 µm s$^{-1}$. In some cases the front reaches the negative electrode and PL is completely removed. It can be later restored by biasing the perovskite film in the contrary direction or letting the system to relax in the absence of bias and dark conditions, although recovery occurs at a much slower rate. These observations where interpreted in terms of ionic redistribution caused by the applied field, but a quantitative understanding in terms of specific ionic species and a mechanistic suppression of PL could not be achieved. In the following, we will present a dynamic transport model, which matches well with experimental data and more importantly directly relates the effects to the drift of halogenide vacancies in perovskites allowing the direct determination of defect densities and diffusion constants.

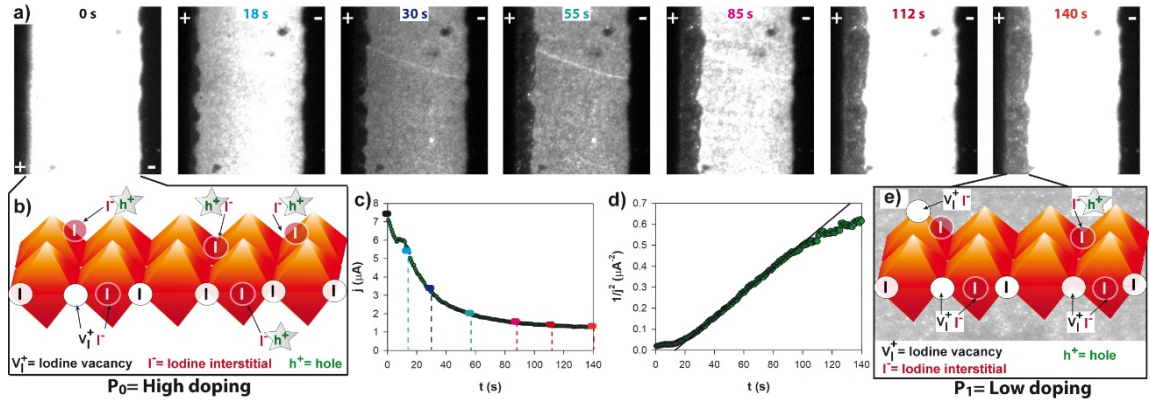

Fig. 1: a) Time dependent PL images of a perovskite film under an external electric field (~5×10⁴ V/m). The '+' and '-' signs indicate the polarity of the electrodes. The excitation intensity is ~35 mW/cm² and the exposure time per image is 200 ms. The channel length is ~150 μm. z(t) represents the PL quenched areas. The time dependent z(t) is displayed in Supporting Information (Figure S1). b) Diagram that highlights the two dominant types of chemical defects present in the crystalline structure of MAPbI₃ that lead to a background doping density: iodine vacancies ($V^+_{I}$) and interstitials ($I^-$). c) and d) Electrical current $j$ and $1/j^2$, respectively, monitored as a function of the time during the measurement of experiment a). e) Diagram that highlights a reduction of doping density in the crystalline structure of MAPbI₃ that leads to a dark front during the PL measurement in a).

In the experiment presented in this work, the electrical current was measured under a bias voltage of 10 V while the PL was optically recorded. In Fig. 1a we observe that the overall initial PL intensity within the channel decreases slightly during the initial 30 seconds. In parallel, the measured current shows a rapid decrease from 7 μA to 3.0 μA after 30 s. Simultaneously, after 18 s a dark front becomes evident in the left hand side of the channel which advances towards the right of the channel during the measurement time. After 85 s this dark front slows down its advance and the current starts to stabilize at about ~2 μA. After this time the dark front advances slowly and after 140 s the front covers ¼ of the channel width. We also note that at this point the PL increases in the bulk of the channel to values similar to those observed initially. We should remark that the observation of a sharp moving front has been confirmed in independent devices that all show this general behavior, see supporting information for a different example (Figure S2 and S3). However, the point at which the dark front slows its advance and current saturates varied for different perovskite material systems, which shows the overall sensitivity to defect densities. In any case there is a clear relationship between the velocity in the advance of the dark front and the measured current.

A variety of effects can be invoked to explain a decrease or increase of PL, however the challenge is to explain the existence of a sharp front that moves at velocity $v$. This effect cannot be caused by removal of electronic carriers swept by the electrical field, that would produce a homogeneous decrease of PL across the film. The low velocity of the sharp front indicates that we must consider the diffusion of ions. However, diffusion of the vacancies would cause a gradually decaying spatial distribution of PL rather than an advancing edge.

In general clarifying the dominant moving species in a mixed ionic-electronic conductor is quite challenging. The work of Senocrate *et al.*[14] combined several techniques including tracer diffusion, stoichiometric variation, conductivity, and polarization experiments, to conclude that MAPbI$_3$ is predominantly p-doped by iodine interstitials (Fig. 1b). Therefore, electronic conductivity is mainly carried by electronic holes. However, the main ionic conductivity component occurs by hopping conductivity of iodine vacancies and not by iodine interstitials. Pointing to this direction, experiments with iodine vapor that increase the concentration of iodine interstitials reduce ionic conductivity. Similarly, experiments replacing Pb$^{2+}$ by Na$^+$ that increase the initial concentration of iodine interstitials do not increase the ionic conductivity. These observations are compatible with the presence of Frenkel defects involving displacement of an atom from the lattice position to an interstitial position and generation of a vacancy. Other works have endorsed the conclusions of these observations[15,16] and the presence of iodine interstitials have recently been measured by X-Ray and neutron scattering.[17]

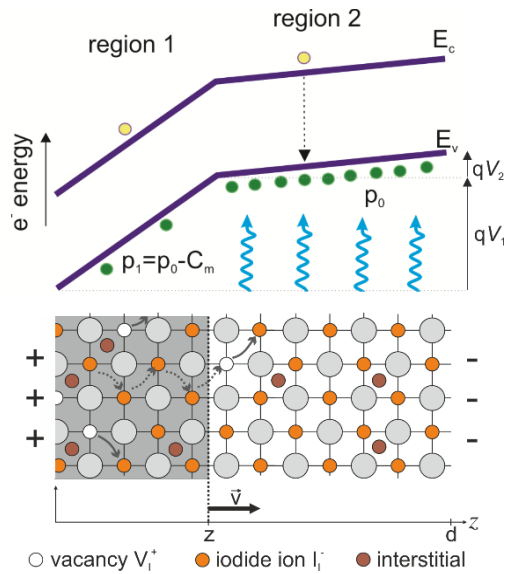

Fig. 2. Scheme of optoelectronic effects of iodine vacancies drifting in the perovskite layer under the applied field caused by positive bias at the left side. Initially the intrinsic doping by iodine interstitials I$_i^-$ creates majority carriers of density $p_0$ that cause PL under photogeneration of electrons as in region 2. With applied voltage $V$ the conduction band $E_c$ and valence band $E_v$ bend an amount q$V$=q$V_1$+q$V_2$, where $q$ is elementary charge. Iodine vacancies V$_I^+$ drift in the electrical field towards the right. They fill the space up to a density $C_m$ in region 1 compensating the p-doping and reducing majority carrier density to $p_1 = p_0 - C_m$. The filling with vacancies up to point $z$ advances with time at a velocity $v$.

Based on the ionic-electronic properties of MAPbI$_3$, we formulate a new dynamic model that is outlined in Fig. 2. As mentioned previously under a constant applied bias iodine vacancies $V_I^+$ drift in the electrical field towards the right. The additional vacancies correspond to iodine ions that accumulate at the contact surface, in a characteristic accumulation region that has been observed in many cases by impedance spectroscopy [18]. When the $V_I^+$ vacancies move to the right they compensate the negative charge from iodine interstitials reducing the hole density as shown in Figure 1e. Vacancies can fill the space up to a maximum concentration value $C_m$, reducing hole density to $p_1 = p_0 - C_m$ (assuming $p_0 > C_m$). The reduction of doping and the creation of nonradiative recombination sites induces a dark region 1 of size $z$ that increases with time. Meanwhile region 2 ($z \leq x \leq d$) remains largely undisturbed with approximately the initial hole density $p_0$.

The electrical current in each region $j_i$ ($i = 1, 2$) is driven by the respective electrical field $E_i$ and since the mobility $\mu_p >> \mu_C$ the ionic current component can be neglected. We have

$$j_1 = q p_1 \mu_p E_1 = j_2 = q p_0 \mu_p E_2 \tag{1}$$

Hence the division of voltage between the two regions in $V = V_1 + V_2$ is governed by the equation

$$V_2 = \frac{p_1}{p_0} \frac{d-z}{z} V_1 \tag{2}$$

and we obtain

$$V_1 = \left[ 1 + \frac{p_1}{p_0} \left( \frac{d}{z} - 1 \right) \right]^{-1} V \tag{3}$$

The ionic drift current in region 1 is

$$j_C = q C_m \mu_C \frac{V_1}{z} \tag{4}$$

The ionic flux causes the increase of region 1 as follows

$$j_C = q C_m \frac{dz}{dt} \tag{5}$$

Combining the previous expressions we obtain the equation for the variation of $z$

$$v = \frac{dz}{dt} = z^{-1} \left[ 1 + \frac{p_1}{p_0} \left( \frac{d}{z} - 1 \right) \right]^{-1} \mu_C V \tag{6}$$

By integration, we get the dependence on time of the velocity of advance of the front

$$v(t) = \left( 1 + \frac{2\gamma}{d} v_0 t \right)^{-1/2} v_0 \tag{7}$$

where the initial velocity is

$$v_0 = \frac{p_0}{p_1} \frac{\mu_C V}{d}$$

(8)

and we have introduced the quantity

$$\gamma = \frac{p_0}{p_1} - 1$$

(9)

The electrical current depends on time as

$$j = \left(1 + \frac{2\gamma}{d} v_0 t\right)^{-1/2} j_0$$

(10)

where the initial current is

$$j_0 = q p_0 \mu_p \frac{V}{d}$$

(11)

Accordingly the following representation of $j(t)$

$$\frac{1}{j^2} = \frac{1}{j_0^{\,2}} + at$$

(12)

forms a straight line up to the point where current saturates. The slope in Equation (12) is given by

$$a = \frac{2\gamma v_0}{d j_0^{\,2}}$$

(13)

The final current at $z = d$ has the value

$$j = \frac{1}{1+\gamma} j_0 = q p_1 \mu_p \frac{V}{d}$$

(14)

It is smaller than the initial current, since the initial amount of p-doping has decreased by the spread of vacancies.

Let us consider alternative scenarios for the modification of defect distribution in the biased film. Instead of a p-doped system assumed above let us consider a system that is n-doped with intrinsic density $n_0$ and undergoes the same experiment. The model is very similar to Fig. 2 but in this case the moving iodine vacancies will reduce the doping of region 1 and this zone will get the larger electrical field. The results are given by the same previous equations but now

$$\gamma = \frac{p_0}{n_1} - 1$$

(15)

Thus the current increases with time, in contrast to the case above, as now the overall doping is increased. Finally we consider the case in which the drifting vacancies change the type of doping so that $n_1 = C_m - p_0$. Then region 1 and 2 conduct by electrons and holes, respectively, and the layer operates in a diode mode by double electronic carrier injection from electrodes. The currents meet at $z$ establishing a strong recombination zone at the edge, as in a light-emitting electrochemical cell (LEC) where the redistribution of ionic charges forms the recombination diode structure.[19] Indeed, some perovskite systems show an increased photoluminescence right at the boundary between region 1 and 2. (Supporting information S4) This indicates that electroluminescence i.e. the direct recombination of holes and electrons occurs at that place and electron conduction occurs as well. Still, in the herein presented model it was completely sufficient to choose a single carrier model of Fig. 2 to describe the experiments.

As it is observed in Fig. 1c that the electrical current decreases with time, we are in the situation of majority hole carriers, where bias-induced excess $V_I^+$ in region 1 reduce the carrier density. Plotting the current as $j^{-2}(t)$ in Fig. 1d an excellent accord is obtained with the model, as the figure displays a straight line in the first half of the length $d = 150$ μm. We obtain the results $j_0 = 5\ \mu A$, $a = 6.0 \times 10^{-3}\ \mu A^{-2}\ s^{-1}$, $v_0 = 5\ \mu m\ s^{-1}$, $\gamma = 2.3$. By Equation (9) it follows that

$$p_1 = \frac{1}{1+\gamma} p_0 = 0.3 p_0$$

(16)

This result indicates a minor decrease of doping density so that the major effect producing the obscure region must be the creation of nonradiative recombination sites. Using Equation (8) we obtain the mobility of iodine vacancies $\mu_C = 2.25$ x $10^{-7}$ cm$^2$V$^{-1}$s$^{-1}$ that corresponds to a diffusion coefficient $D_{V_I} = 6 \times 10^{-9}$ cm$^2$s$^{-1}$ in good agreement with the value $2 \times 10^{-9}$ cm$^2$s$^{-1}$ of Senocrate et al.[14] Beyond 100 sec we observe a deviation from the modelled curve and the measured slows down to decrease. Region 2 cannot be assumed to remain undisturbed with $p_0$ at this point.

The presence of several types of defects have been theoretically predicted in MAPbI$_3$, those include methyl ammonium interstitials, Pb vacancies and Pb interstitials[16]. However, in the current work we describe a simple model that takes into account only iodine interstitials and iodine vacancies. This model is able to explain the suppression of luminescence and the existence of a sharp front. The insight allows us to consider in detail the complex dynamics of transport of the defects and how they affect semiconductor properties in the perovskite layer.[14] It is important to highlight that iodine interstitials have been measured to be in very high concentration at room temperature.[17] The mechanism of generation of these interstitial defects in not clear, however, generation of Frenkel defects by the displacement of an atom from the lattice position is a possibility with very low activation energies of 0.06 eV.[20] In this situation, an interstitial iodine and a iodine vacancy is generated and these two species are the only two species required in our model. Once generated the iodine vacancy, activation energies for iodine transport from interstitial have been calculated to be below 0.1 eV by DFT.[21]

In conclusion, an electrical bias in a perovskite layer can switch on and off the photoluminescence in a time scale of seconds. We explain the advance of the dark area with a sharp front in terms of the drift of iodine vacancies that fill the space up to a critical density turning the material more intrinsic-like and enhancing nonradiative recombination. An interplay between interstitial and vacancy defects determines the dominant electronic density and subsequently allows a control of electro-optical properties.

The model fits perfectly with the experimental data describing the initial dynamics of the drift of vacancies in an electric field and directly allows to determine the mobility of iodine vacancies. This method provides a direct and visual way to track the migration of vacancies and obtain a value for the defect density. It is applicable to a broad range of organometal halide perovskite systems which will greatly help to understand how to compensate defect and migration processes in order to eliminate hysteresis effects and device degradation.

## Acknowledgments

We acknowledge funding from MINECO of Spain under Project MAT2016-76892-C3-1-R. A. G. would like to thank the Spanish Ministerio de Economía y Competitividad for a Ramón y Cajal Fellowship (RYC-2014-16809). C. L, and S. H gratefully acknowledge the financial support by the Bavarian State Ministry of Science, Research, and the Arts for the Collaborative Research Network ''Solar Technologies go Hybrid''.

## Experimental Section.

**Device Fabrication.** $CH_3NH_3PbI_3$ precursor solution was prepared by dissolving 0.88 M lead chloride ($PbCl_2$) (Sigma-Aldrich) and 2.64 M $CH_3NH_3I$ (MAI) (Tokyo Chemical Industry company) in anhydrous N,N-dimethylformamide (DMF) (99.8%, Sigma-Aldrich). The precursor solutions were filtered through a 0.2 μm polytetrafluoroethylene (PTFE) filter (Carl Roth GmbH+Co. KG). Glass substrates were washed with acetone, and isopropanol for 10 min each. Then these glass substrates were treated with Ozone for around 10 min. In a nitrogen gas filled glovebox, this precursor solution was spin-coated on the glass substrates at 3000 rpm for 60 s. Following that, the as-spun films were annealed at 100 °C for 80 min in the glovebox. The perovskite film on glasses were transferred into an evaporation chamber with pressure of ~3×10$^{-6}$ mbar, and ~70 nm thickness of Au was deposited by thermal evaporation through a shadow mask. This interdigitating shadow mask defined the electrode geometry: the electrode distance was 200 μm and a ratio between channel width W and length L, W/L of 500. External conducting wires were connected to the device using an Ultrasonic Soldering System (USS-9200, MBR electronics GmbH). In the end, to protect the film from oxygen and water, 40 mg/mL poly(methyl methacrylate) (PMMA) solution dissolved in butyl acetate (anhydrous, 99%, Sigma-Aldrich) was spin-coated on the film at a speed of 2000 rpm for 60 sec in the glovebox.

**PL Imaging Microscopy**. This measurement was conducted on a home-build PL microscope, as shown in Figure S5. Based on a commercial microscopy (Microscope Axio Imager.A2m, Zeiss), the sample was allocated in the focal plane of an objective lens (10×/0.25 HD, Zeiss), and the sample position was adjusted by a motorized scanning stage (EK 75*50, Märzhäuser Wetzlar GmbH & Co. KG). The sample was illuminated by an internal LED illuminator using a filter (HC 440 SP, AHF analysentechnik AG) with the excited wavelength of around 440 nm. The excitation light intensity can be controlled and was set to ~35 mW/cm$^2$. The PL signal was filtered (HC-BS 484, AHF analysentechnik AG) to suppress residual excited light and directed onto a CCD camera (Pco. Pixelfly, PCO AG) with the exposure time of 200 ms. A constant voltage of 10 V was applied between the Au electrodes (236 Source Measure Unit, Keithley Company), and the current was monitored and recorded by a LabVIEW program.

## Supporting Information

Time dependent growth of the dark region; experimental behavior of different samples; spatial profile of PL intensity within the channel; diagram for a PL imaging microscopy.

## Contributions

C.L., A.G. and S.H. made the experiments, J.B. made the model, C.L., A.G., S.H. made the interpretation of experimental data and wrote the paper.

## References

(1)      Panzer, F.; Li, C.; Meier, T.; Köhler, A.; Huettner, S. Impact of Structural Dynamics on the Optical Properties of Methylammonium Lead Iodide Perovskites, *Adv. Energy Mater.* **2017**, *7*, 1700286-n/a.

(2)      Lopez-Varo, P.; Jiménez-Tejada, J. A.; García-Rosell, M.; Ravishankar, S.; Garcia-Belmonte, G.; Bisquert, J.; Almora, O. Device Physics of Hybrid Perovskite Solar cells: Theory and Experiment, *Adv. Energy Mater.* **2018**, 1702772.

(3)      Sanchez, R. S.; Gonzalez-Pedro, V.; Lee, J.-W.; Park, N.-G.; Kang, Y. S.; Mora-Sero, I.; Bisquert, J. Slow dynamic processes in lead halide perovskite solar cells. Characteristic times and hysteresis, *J. Phys. Chem. Lett.* **2014**, *5*, 2357−2363.

(4)      Hoke, E. T.; Slotcavage, D. J.; Dohner, E. R.; Bowring, A. R.; Karunadasa, H. I.; McGehee, M. D. Reversible photo-induced trap formation in mixed-halide hybrid perovskites for photovoltaics, *Chemical Science* **2015**, *6*, 613-617.

(5)      Leijtens, T.; Srimath Kandada, A. R.; Eperon, G. E.; Grancini, G.; D'Innocenzo, V.; Ball, J. M.; Stranks, S. D.; Snaith, H. J.; Petrozza, A. Modulating the Electron–Hole Interaction in a Hybrid Lead Halide Perovskite with an Electric Field, *J. Am. Chem. Soc.* **2015**, *137*, 15451-15459.

(6)     Greenham, N. C.; Peng, X.; Alivisatos, A. P. Charge separation and transport in conjugated-polymer/semiconductor-nanocrystal composites studied by photoluminescence quenching and photoconductivity, *Phys. Rev. B* **1996**, *54*, 17628-17637.

(7)     Morana, M.; Wegscheider, M.; Bonanni, A.; Kopidakis, N.; Shaheen, S.; Scharber, M.; Zhu, Z.; Waller, D.; Gaudiana, R.; Brabec, C. Bipolar Charge Transport in PCPDTBT-PCBM Bulk-Heterojunctions for Photovoltaic Applications, *Adv. Func. Mater.* **2008**, *18*, 1757-1766.

(8)     Galisteo-López, J. F.; Li, Y.; Míguez, H. Three-Dimensional Optical Tomography and Correlated Elemental Analysis of Hybrid Perovskite Microstructures: An Insight into Defect-Related Lattice Distortion and Photoinduced Ion Migration, *J. Phys. Chem. Lett.* **2016**, *7*, 5227-5234.

(9)     Yoon, S. J.; Kuno, M.; Kamat, P. V. Shift Happens. How Halide Ion Defects Influence Photoinduced Segregation in Mixed Halide Perovskites, *ACS Energy Lett.* **2017**, *2*, 1507-1514.

(10)     Jacobs, D. L.; Scarpulla, M. A.; Wang, C.; Bunes, B. R.; Zang, L. Voltage-Induced Transients in Methylammonium Lead Triiodide Probed by Dynamic Photoluminescence Spectroscopy, *J. Phys. Chem. C* **2016**, *120*, 7893-7902.

(11)     Xu, Z.; De Rosia, T.; Weeks, K. Photoluminescence–Voltage (PL–V) Hysteresis of Perovskite Solar Cells, *J. Phys. Chem. C* **2017**, *121*, 24389-24396.

(12)     Chen, S.; Wen, X.; Huang, S.; Huang, F.; Cheng, Y.-B.; Green, M.; Ho-Baillie, A. Light Illumination Induced Photoluminescence Enhancement and Quenching in Lead Halide Perovskite, *Solar RRL* **2017**, *1*, 1600001-n/a.

(13)     Li, C.; Guerrero, A.; Zhong, Y.; Gräser, A.; Luna, C. A. M.; Köhler, J.; Bisquert, J.; Hildner, R.; Hüttner, S. Real-Time Observation of Iodide Ion Migration in Methylammonium Lead Halide Perovskites, *Small* **2017**, 1701711.

(14)     Senocrate, A.; Moudrakovski, I.; Kim, G. Y.; Yang, T.-Y.; Gregori, G.; Grätzel, M.; Maier, J. The Nature of Ion Conduction in Methylammonium Lead Iodide: A Multimethod Approach, *Angew. Chem. Int. Ed.* **2017**, *56*, 7755-7759.

(15)     Wei, L.-y.; Ma, W.; Lian, C.; Meng, S. Benign Interfacial Iodine Vacancies in Perovskite Solar Cells, *J. Phys. Chem. C* **2017**, *121*, 5905-5913.

(16)     Meggiolaro, D.; Mosconi, E.; De Angelis, F. Modeling the Interaction of Molecular Iodine with MAPbI3: A Probe of Lead-Halide Perovskites Defect Chemistry, *ACS Energy Lett.* **2018**, *3*, 447-451.

(17)     Minns, J. L.; Zajdel, P.; Chernyshov, D.; van Beek, W.; Green, M. A.

Structure and interstitial iodide migration in hybrid perovskite methylammonium lead iodide, *Nat. Commun.* **2017**, *8*, 15152.

(18)     Zarazua, I.; Bisquert, J.; Garcia-Belmonte, G. Light-induced space-charge accumulation zone as photovoltaic mechanism in perovskite solar cells, *J. Phys. Chem. Lett.* **2016**, *7*, 525-528.

(19)     Lenes, M.; Garcia-Belmonte, G.; Tordera, D.; Pertegás, A.; Bisquert, J.; Bolink, H. J. Operating Modes of Sandwiched Light-Emitting Electrochemical Cells, *Adv. Func. Mater.* **2011**, *21*, 1581-1586.

(20)     Son, D.-Y.; Kim, S.-G.; Seo, J.-Y.; Lee, S.-H.; Shin, H.; Lee, D.; Park, N.-G. Universal Approach toward Hysteresis−Free Perovskite Solar Cell via Defect Engineering, *J. Am. Chem. Soc.* **2018**.

(21)     Azpiroz, J. M.; Mosconi, E.; Bisquert, J.; De Angelis, F. Defect migration in methylammonium lead iodide and its role in perovskite solar cell operation, *Energy Environ. Sci.* **2015**, *8*, 2118--2127.

**Supporting Information**

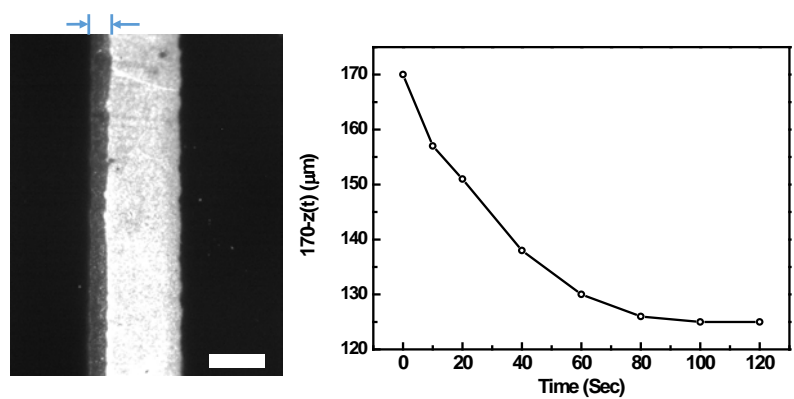

Figure S1. a) PL image of a perovskite film under an external electric field (~5×10$^4$ V/m). z(t) represents the PL quenched area. b) The time dependent z(t) is shown. This time dependent curve is consistent with the measured time dependent current.

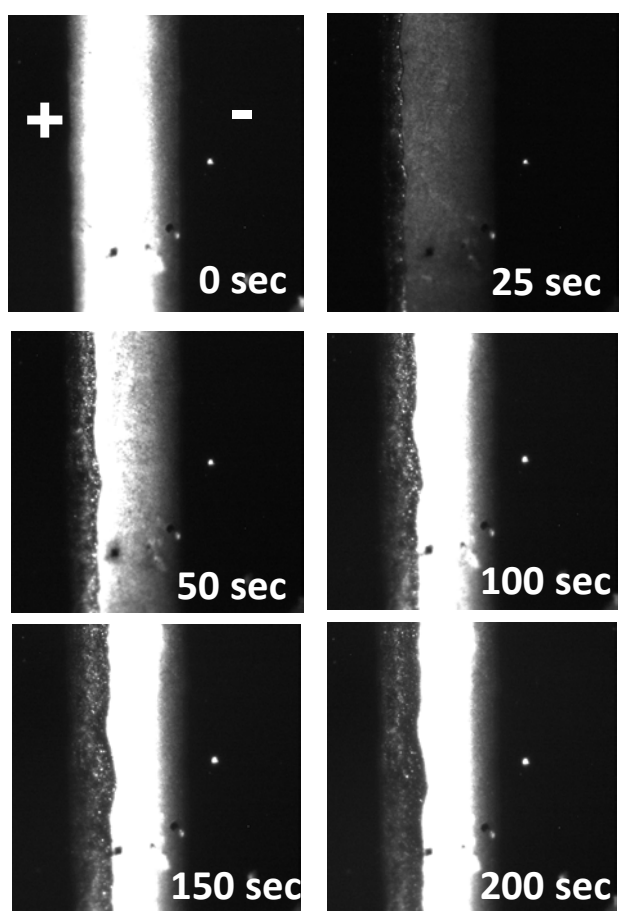

Figure S2. Time dependent PL images of a perovskite film under an external electrical field (~1.3×10$^5$ V/m). The '+' and '-' signs indicate the polarity of the electrodes. The excitation intensity is ~35 mW/cm$^2$ and the exposure time per image is 200 ms. The channel length is ~150 µm. With a larger electric field, the PL inactive area expands more compared with the samples with smaller field in Fig 1a.

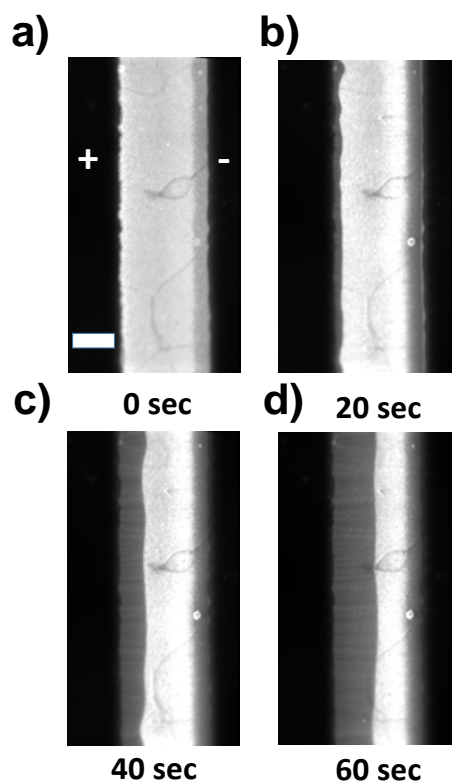

a)  b)  0 sec
c)  20 sec  d)  40 sec  60 sec

Figure S3. Time dependent PL images of a 2D multiple quantum wells perovskite film (NFPI$_7$) *(1)* under an external electrical field (~3×10$^5$ V/m). The '+' and '-' signs indicate the polarity of the electrodes. The excitation intensity is ~35 mW/cm$^2$ and the exposure time per image is 200 ms. The scale bar is 100 µm.

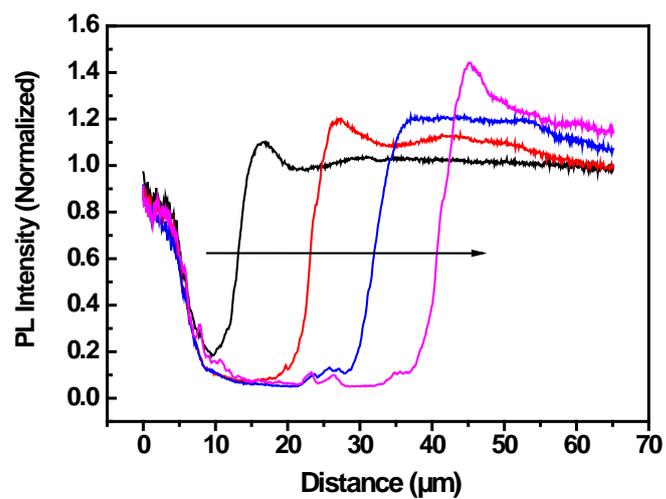

Figure S4: Integrated and normalized PL intensity showing the PL profile within the channel. The quenched PL area is moving from left to right. The PL intensity increases through electroluminescence right at the border between region 1 and region 2.

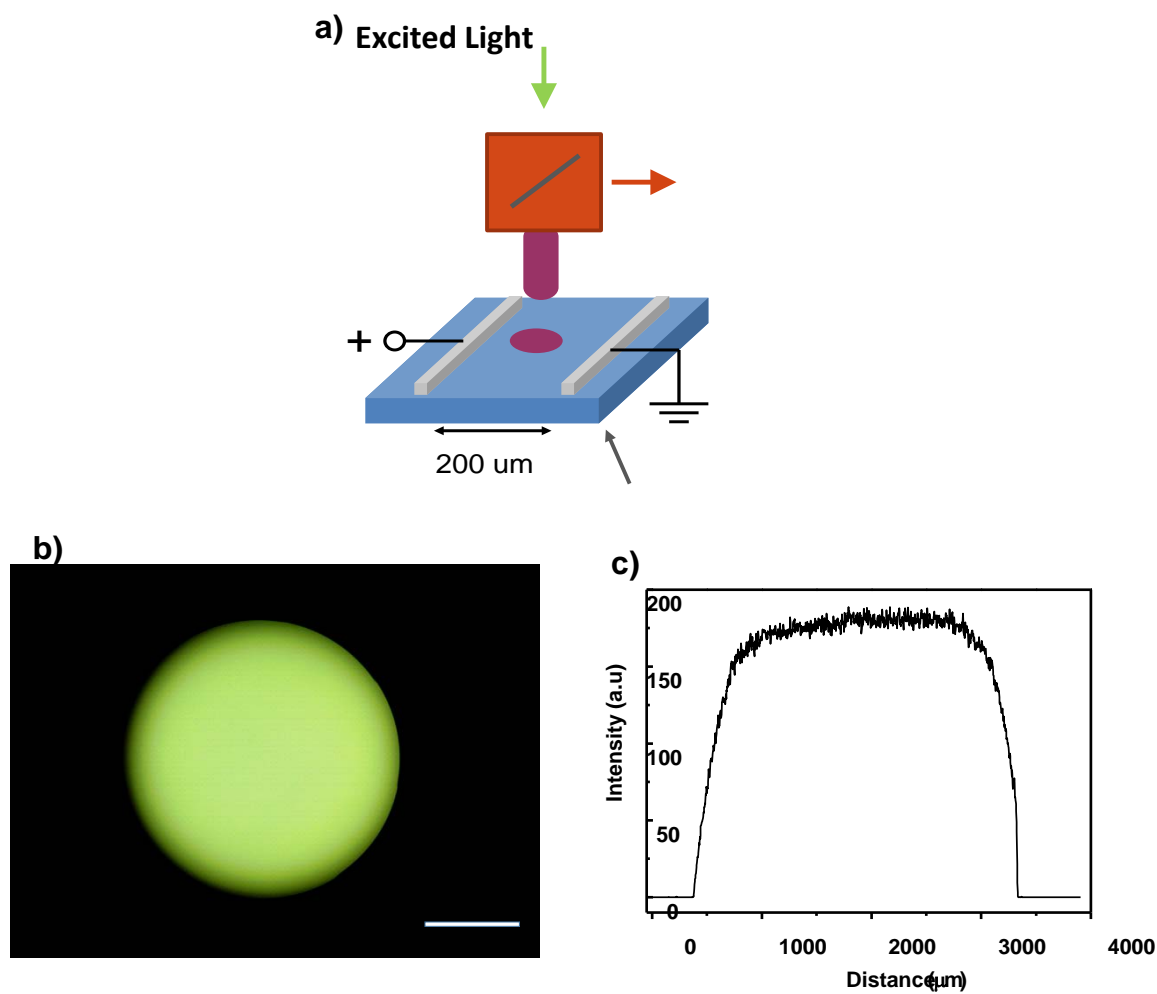

Figure S5. a) Schematic diagram for a PL imaging microscopy. b) Excited light beam (wavelength ~440 nm) image captured by a CCD camera. The scale bar is 1 mm. c) Intensity distribution of the excited beam in the focus plane. Based on this intensity profile, we can consider that at the central of the beam, it is a uniformly distributed light intensity, ~35 mW/cm$^2$.

**References:**

1. N. Wang *et al,* Perovskite light-emitting diodes based on solution-processed self-organized multiple quantum wells, *Nat. Photon.***10**, 699 (2016).

2. C. Li *et al.*, Real-Time Observation of Iodide Ion Migration in Methylammonium Lead Halide Perovskites. *Small*, ***13,*** 1701711 (2017).